\documentstyle[11pt,aaspp4,epsf]{article}

\begin{document}

\newcommand{\ci}{[\ion{C}{1}]}
\newcommand{\twcoft}{$^{12}$CO $J$=4$-$3}
\newcommand{\twcott}{$^{12}$CO $J$=3$-$2}
\newcommand{\twcoto}{$^{12}$CO $J$=2$-$1}
\newcommand{\twcooz}{$^{12}$CO $J$=1$-$0}
\newcommand{\twco}{$^{12}$CO}
\newcommand{\thcoft}{$^{13}$CO $J$=4$-$3}
\newcommand{\thcott}{$^{13}$CO $J$=3$-$2}
\newcommand{\thcoto}{$^{13}$CO $J$=2$-$1}
\newcommand{\thco}{$^{13}$CO}
\newcommand{\jft}{$J$=4$-$3}
\newcommand{\jtt}{$J$=3$-$2}
\newcommand{\jto}{$J$=2$-$1}
\newcommand{\joz}{$J$=1$-$0}
\newcommand{\etal}{et al.}
\newcommand{\eg}{e.g.}
\newcommand{\ie}{i.e.}
\newcommand{\cf}{cf.}
\newcommand{\kms}{km s$^{-1}$}
\newcommand{\tas}{$T_{\rm A}^{\star}$}
\newcommand{\trs}{$T_{\rm R}^{\star}$}
\newcommand{\tmb}{$T_{\rm MB}$}
\newcommand\itci{C{\small\it{I}}\relax}

\title{The Physical Conditions and Dynamics of the Interstellar Medium
in the Nucleus of M83: Observations of CO and \ion{C}{1} } 
\author{Glen R. Petitpas and Christine D. Wilson}
\affil{McMaster University, 1280 Main Street West, Hamilton Ontario,
Canada L8S 4M1}
\authoremail{petitpa@impatiens.physics.mcmaster.ca}

\begin{abstract}

This paper presents \ci, \twcoft, and \twcott\ maps of the barred
spiral galaxy M83 taken at the James Clerk Maxwell Telescope.
Observations indicate a double peaked structure which is consistent
with gas inflow along the bar collecting at the inner Lindblad
resonance. This structure suggests that nuclear starbursts can occur
even in galaxies where this inflow/collection occurs, in contrast to
previous studies of barred spiral galaxies. However, the observations
also suggest that the double peaked emission may be the result of a
rotating molecular ring oriented nearly perpendicular to the main
disk of the galaxy. The \twcoft\ data indicate the presence of warm gas
in the nucleus that is not apparent in the lower $J$ CO observations,
which suggests that \twcooz\ emission may not be a reliable tracer of
molecular gas in starburst galaxies. The twelve \ci/\twcoft\ line
ratios in the inner 24$''\times 24''$ are uniform at the 2$\sigma$
level, which indicates that the \twcoft\ emission is originating in the
same hot photon-dominated regions as the \ci\ emission. The
\twcoft/\jtt\ line ratios vary significantly within the nucleus with the
higher line ratios occurring away from peaks of emission along an arc
of active star forming regions. These high line ratios ($>$1) likely
indicate optically thin gas created by the high temperatures caused by
star forming regions in the nucleus of this starburst galaxy.

\end{abstract}


\section{Introduction}

The circumnuclear regions of galaxies are very often the setting for
starbursts and other extraordinary events. Observations of the gas
kinematics and distribution indicate that bars, resonances, gas inflow,
and tidal shear play important roles in the formation and evolution of
nuclear starbursts (\eg\ Handa \etal\ 1990; Kenney \etal\ 1992).
Previous observations show that the molecular gas in the central
regions of barred spiral galaxies often does not extend all the way
into the centre of the nucleus. It accumulates some distance away from
the centre, giving the emission a double peaked appearance with each
peak occurring where the bar meets the nucleus (Kenney \etal\ 1992;
Ishizuki \etal\ 1990). Dynamical models indicate that in the presence
of a barred potential, gas will flow inward along the bar and slow its
descent temporarily near inner Lindblad resonances (ILR, \eg\
Combes 1988; Shlosman, Frank \& Begelman 1989). At these locations the
gas may accumulate into larger complexes of molecular
clouds.

In addition to complex dynamics, there is most likely a profusion of
complicated photo-chemistry occurring within the nuclei of starburst
galaxies. Interstellar clouds are believed to consist of smaller high
density dark cores interspersed throughout a larger region of lower
density. These high density regions are self-shielded from ultraviolet
(UV) radiation that tends to dissociate molecular gas. The result is
the population of the diffuse region by hydrogen atoms (\ion{H}{1}),
atomic and ionized carbon (C, C$^+$), and many other atomic and ionized
species (\eg\ Morton \etal\ 1973), while the dark cores can contain
molecular species such as H$_2$ and CO.

It is believed that atomic carbon can exist only in a small energy
window, outside of which it will either be ionized or combined with
oxygen to form CO. Inside the dense cores, most of the carbon combines
to form CO, while outside the cores, the UV radiation acts to ionize
atomic carbon. It is therefore expected that \ion{C}{1} is the dominant
species near the edges of dense, self-shielding cloud cores.

This simple model fails to explain the extended \ci\ emission observed
in molecular clouds in our own Galaxy (Plume, Jaffe, \& Keene 1994;
Keene \etal\ 1985). One possible explanation is found in the clumpy
structure of molecular clouds (\eg\ Stutzki \& G\"usten 1990). This
clumpiness would allow UV radiation to penetrate much deeper into the
cloud allowing atomic carbon to exist at depths greater than would be
allowed by a simple spherical model of molecular clouds (\eg\ Boisse
1990). Many alternative explanations have been proposed to explain the
extended \ci\ emission. These ideas range from complicated chemical
processes involving H$^+$ (Leung, Herbst, \& Heubner 1984) to simpler
ideas such as a C/O ratio greater than one (\eg\ Keene \etal\ 1985).

This paper presents \ci\ and CO maps of the barred spiral galaxy M83.
Its low inclination angle ($i$ = 24\arcdeg, Comte 1981) and close
proximity ($D$ = 4.7 Mpc, Tully 1988) make it one of the best locations
for studying the response of gas to a barred potential. It is believed
to be undergoing a nuclear starburst (\eg\ Talbot \etal\ 1979), which
may have been triggered by molecular gas inflow along the bar
potential. This nuclear starburst would produce higher temperatures
which should readily excite the higher $J$-transitions in the CO gas.
Also, strong UV flux has been detected in the nucleus (Bohlin
\etal\ 1983), which would help dissociate the CO into atomic carbon. By
studying the CO and \ci\ data, we can understand better the dynamics of
the gas and its role in fueling the nuclear starburst and also learn
about the conditions conducive to the formation of atomic carbon.

\section{Observations and Data Reduction \label{obs}}

\ci, \twcoft, and \twcott\ observations of M83 were taken using the 15
m James Clerk Maxwell Telescope (JCMT) over the period of 1996 April 5
-- 10. The half-power beam width of the JCMT is $12''$ at 492 GHz (\ci)
and 461 GHz (\twcoft) and $14''$ at 345 GHz (\twcott). The galaxy was
mapped using 5$''$ sampling in all transitions. All observations
were obtained using the Dwingeloo Autocorrelation Spectrometer.

The data were reduced using the data reduction package SPECX and were
binned to 4 km s$^{-1}$ resolution. The raw spectra were converted into
FITS files using SPECX and then the Bell Labs data reduction package
COMB was used to convolve the \twcoft\ to the same beam size as the
\twcott\ data. The \twcott\ data were taken in raster mode at the JCMT.
It was later discovered that a loose wire was causing poor baselines
for spectra taken in this manner so the \twcoft\ and \ci\ data were
taken in beam-switched mode. The beam was switched to a position
60\arcsec\ perpendicular to the bar, which we assume to have negligible
emission with respect to the on-positions located along the galactic
bar. The data were binned and zeroth order baselines were removed. The
poorer baselines encountered for the \twcott\ raster maps resulted in
poor cosmetic appearance of the intensity maps in the regions without
emission, but did not noticeably effect the line parameters. Pointing
was checked frequently and was typically less than 2\arcsec\ rms. The
calibration was monitored by frequently observing both planets and
spectral line calibrators. The spectral line calibrators for the CO
data had integrated intensities that were within $\sim$10\% of the
published values. We therefore adopt the values of $\eta_{\rm MB}$ =
0.58 for \twcott\ and $\eta_{\rm MB}$ = 0.52 for the \twcoft\ and
\ci\ data as given in the JCMT User's Guide.

Since the line strengths we are comparing are measured with (or
convolved to) the same beam diameter, using the \trs\ temperature scale
would ensure that our observed line ratios are equal to the true
radiation temperature ratios. However, conversion to \trs\ from \tas\
requires knowledge of the forward scattering and spillover ($\eta_{\rm
FSS}$), which is difficult to measure and was not attempted during the
observing run. We do have good values for the main beam efficiencies
and so an accurate conversion to main beam temperature (\tmb) is
possible. Using the \tmb\ scale instead of \trs\ scale only changes the
\twcoft/\jtt\ line ratio by $\sim$10\% under normal calibration
conditions. For this reason, we will use the main beam temperature
scale for the line ratios throughout this paper. The maps are also plotted
in the \tmb\ temperature scale. The individual \twcott, \twcoft, and
\ci\ spectra are shown in Figures \ref{32grsp}, \ref{43grsp}, and
\ref{CIgrsp} respectively.

\section{Morphology and Dynamics}

The integrated intensity maps in the three lines (Figures
\ref{32map}--\ref{CImap}) indicate a double (possibly triple) peaked
structure in the emission. Within the pointing uncertainty of the JCMT
(2\arcsec\ rms), each map exhibits the strongest emission peak at
(+5\arcsec, +5\arcsec) with the second strongest at ($-$5\arcsec,
$-$10\arcsec). These two emitting regions peak at different
velocities: the channel maps (Figures
\ref{32chanmap}--\ref{43chanmap}) show that the emission at (+5\arcsec,
+5\arcsec) peaks at approximately 500 \kms\ in both the \twcott\ and
\twcoft\ maps, while the emission at ($-$5\arcsec, $-$10\arcsec) peaks
at $\sim$550 \kms. The \ci\ emission is too weak to obtain reliable
channel maps. In the integrated intensity maps (especially \ci\ and
\twcoft), there are indications of a third, weaker peak at
approximately ($-$10\arcsec, $-$5\arcsec), but the channel maps do not
indicate a separate clump of emission at a velocity different from the
two main emission peaks. The channel maps suggest that the hints of a
third emission peak are likely just an extension of the ($-$5\arcsec,
$-$10\arcsec) peak.

\subsection{Gas Collection at Lindblad Resonances?}

Figures \ref{32map}, \ref{43map}, and \ref{CImap} clearly show that the
\ci\ emission appears to follow the \twcoft\ emission much more closely
than it does the \twcott\ emission. The \twcoft\ and \ci\ emission
appear asymmetric, with the two main peaks to the north and south of
the centre connected by an ``elbow'' of weaker emission that undergoes
a bend of $\sim$90\arcdeg. Irregular structure was also seen in M83 in
the \twcooz\ emission by Handa \etal\ (1990). Numerical simulations
predict that a shock front will form at the leading edge of the bar
(\eg\ S\o rensen \etal\ 1976). At this shock front, interstellar gas is
compressed and loses kinetic energy, which causes it to flow toward the
nucleus. The star formation along the bar is thought to be triggered by
such an inflow. In the figures, where the galactic bar of M83 runs
along the $x$-axis, it is clear that the emission is originating on the
leading edge of the bar potential as predicted by theoretical models of
barred galaxies. This geometry in M83 was first identified by Handa
\etal\ (1990).

One important question is why the gas does not continue to flow into
the centre of the galaxy. It has been suggested (\eg\ Kenney
\etal\ 1992) that the location of the ILR is often coincident with the
double peaks of molecular gas often seen in barred spiral galaxies.
They studied several barred spiral galaxies and found that galaxies
without nuclear starbursts exhibit a double peaked emission structure,
while the galaxy with the nuclear starburst has a single peak of
emission at the center. This result suggests that accumulation of gas
at ILRs is a way of inhibiting nuclear starbursts in barred spiral
galaxies. Moreover, as calculations indicate the starburst galaxy does
contain ILRs, the mere existence of ILRs does not appear to be
sufficient to prevent starbursts (Kenney \etal\ 1992).

Although we cannot directly calculate the locations of the ILRs from
our data or existing data (although see Gallais \etal\ (1994) for a
good attempt), the double peaked structure we see in M83 is very
similar to that observed in the non-starburst galaxies of Kenney
\etal\ (1992) and most likely indicates gas collection at an ILR. We
also note that M83 is undergoing a nuclear starburst (\eg\ Talbot
\etal\ 1979) which suggests that {\it a nuclear starburst may occur
even in galaxies that exhibit this double peaked structure}, in
contrast to the findings of Kenney \etal\ (1992). A better rotation
curve for the nucleus of M83 will provide the true location on the
ILRs, thus proving or disproving the above result.

High resolution \twcooz\ maps by Handa, Ishizuki, \& Kawabe (1994) show
that the double peaked structure of M83 straddles a nuclear star
formation complex traced by radio continuum emission (Cowen \& Branch
1985). They do not see strong \twcooz\ emission from within the central
star forming region itself. Our \twcott\ data show a very similar
structure (Figure \ref{32chanmap}, $V_{\rm LSR}$ = 520 \kms).
Moreover, the \twcoft\ data (Figure \ref{43chanmap}) show a very
different structure; at 520 \kms, the \twcoft\ emission {\it peaks} at
the centre of the star forming region. These data suggest that there
{\it is} molecular gas in the nuclear regions of M83, but it is heated
by the nuclear starburst so that most of the CO is collisionally
excited to the higher rotational ($J$) states. This result, combined
with the double peaked accumulations of \twcoft\ emission seen in the
integrated intensity maps (Figure \ref{43map}), suggests that molecular
gas gathers at the ILR before eventually flowing into the nucleus where
the starburst occurs. The gas at the ILRs would be heated less
intensely by the starburst than the gas in the nucleus, so in the
double peaks we see emission from cooler gas (as traced by
\twcooz\ emission) as well as hotter gas (as traced by \twcott\ and
\twcoft\ emission), while in the nucleus most of the gas is in the $J$
= 4 and higher states. This result indicates that \twcooz\ emission is
not always a good tracer of the molecular gas in starburst galaxies, as
is often assumed. The edge-on starburst galaxy M82 exhibits a similar
structure with the \twcoft\ data filling in the torus of rotating gas
traced by the \twcott\ emission (White \etal\ 1994; Tilanus
\etal\ 1991).

\subsection{Rotating Gas Disk/Torus Out of the Plane of the Galaxy?} 

An alternate explanation for the observed CO morphology and kinematics
in M83 is the existence of a rotating torus of molecular gas (\cf\ M82,
Tilanus \etal\ 1991, NGC 253, Israel \etal\ 1995; both also starburst
galaxies). However, the observed \twcoft\ emission that peaks at the
centre of the nucleus is not consistent with a simple torus model.
Nevertheless, the nuclear \twcoft\ emission could be explained if the
CO gas is actually a disk/torus combination. The nuclear
\twcoft\ emission could originate in a disk of hotter molecular gas
located inside a molecular torus of gas. In this picture, the torus
would consist of slightly cooler gas that allows for the presence of
the lower $J$ states.

It has been proposed (\eg\ Larkin \etal\ 1994) that a torus of gas may
accumulate in the regions between the inner and outer ILRs. While this
may be occurring in M83, the problem is that the morphology requires
the torus/disk to be seen nearly edge-on. Since M83 has an inclination
angle of only 24\arcdeg, only a torus lying out of the plane of the
galaxy could possibly produce the observed morphology. However, it is
not entirely unreasonable that there could be a torus/disk of gas that
lies out of the plane of the galaxy. Kinematically distinct subsystems
have been found in at least three disk galaxies (NGC 3672, NGC 4826,
and NGC 253, see Anantharamaiah \& Goss 1996). This configuration may
be more common than the observations indicate due to a selection effect
caused by the high levels of extinction in the nuclear regions of disk
galaxies. M83 has long been believed to have a disk that is quite
warped (Rogstad, Lockhart, \& Wright 1974), while more recent
observations of dust lanes suggest that M83 is undergoing
three-dimensional accretion in the nucleus and that the nuclear dust
lanes pass out of the plane of the galaxy (Sofue \& Wakamatsu 1994).
This model would be generally consistent with the idea of a tilted
torus of molecular gas. Dust tends to trace regions of high density,
while molecules need higher densities to be shielded from dissociation
by UV flux. Hence, it is reasonable to expect a ring of molecular gas
to be coincident with the ring formed by the warped dust lanes that lie
out of the plane of the galaxy.

One problem with the above discussion is explaining {\it how} the inner
part of the disk of M83 could become misaligned with the rest of the
disk of the galaxy. One possible explanation is that M83 has accreted a
smaller companion galaxy. Simulations of disk galaxies accreting a
smaller companion have shown that these `minor mergers' are able to
produce warped disks, large spiral arms, and radial inflow of gas that
may trigger nuclear starbursts (\eg\ Quinn \etal\ 1993; Mihos \&
Hernquist 1994; Hernquist \& Mihos 1995). If the companion was small
enough, it may remain intact as it settles into the nucleus of the
larger galaxy (\eg\ Balcells \& Quinn 1990). The rotation curve that
would be created by the combination of the two rotating objects would
appear very peculiar. It may manifest itself as a discontinuity at the
radius of the smaller galaxy that has been accreted to the core of the
larger galaxy. Such a rotation curve has been predicted by the model of
Handa \etal\ (1990) who fitted the {\it outer} rotation curve of M83 in
an attempt to determine the rotation curve of the inner regions. If an
accretion event is the cause of the proposed inner disk misalignment,
it is not clear whether such a system would have Lindblad resonances
where gas could collect. Better models of mergers of barred galaxies as
well as observation of the inner rotation curve for M83 will yield a
clearer picture of the dynamics at work.

\section{Physical Conditions}

Determining the physical conditions of molecular clouds in other
galaxies can be very difficult. For single dish observations, the beam
size is usually larger than the angular diameter of typical molecular
clouds, and so the line strengths are beam-diluted. Molecular clouds
are known to be very clumpy (\eg\ Stutzki \& G\"usten 1990), and so the
average emission likely originates in a mixture of high and low density
material. In addition, assumptions often have to made about the optical
depth of the emission lines. In the discussion below, we will initially
assume that the \twcoft\ and rarer \ci\ emission are optically thin
($\tau \ll 1$). In addition, we will concentrate on the larger peak
located at (+5\arcsec, +5\arcsec) for which we have strong emission and
good signal-to-noise ratios for all spectra including \ci.

\subsection{Column Densities from Line Intensities}

Atomic carbon emission is rather weak compared to CO emission
(\eg\ White \etal\ 1994; Israel \etal\ 1995) and is often thought to be
optically thin. In the low optical depth limit, the column density is
given by
$$N({\rm C}) = 1.9 \times 10^{15}(e^{E_1/kT_{\rm ex}} + 3 + 5e^{(E_1
-E_2)/kT_{\rm ex}})\int T_{\rm MB} dv ~~~({\rm cm^{-2}}) $$
(Schilke \etal\ 1993; Phillips \& Huggins 1981), where $E_1/k$ is the
energy of the $J$ = 1 level (23.6 K), $E_2/k$ is the energy of the
$J$=2 level (62.5 K), $T_{\rm ex}$ is the excitation temperature of the
gas, and $\int T_{\rm MB} dv$ is the integrated intensity of the
\ci\ line.

If we assume an excitation temperature of 24 K in M83 (Wall \etal\
1993), the strength of the \ci\ line at the (+5\arcsec,+5\arcsec)
emission peak of 71.3 K \kms\ (\tmb) gives an atomic carbon column
density of $N$(C) = $1 \times 10^{18}$ cm$^{-2}$. (Adopting a higher
excitation temperature of 50 K increases the column density by only
15\%.) This column density is somewhat lower than that found in other
starburst galaxies. For example, Israel \etal\ (1995) find atomic
carbon column densities of $7.7 \times 10^{18}$ cm$^{-2}$ in NGC 253
and White \etal\ (1994) find $N$(C) = $1.9 \times 10^{18}$ cm$^{-2}$ in
the centre of M82. The larger column densities seen in both NGC 253 and
M82 may be due to their nearly edge-on orientation ($i$ = 78\fdg 5 for
NGC 253, Pence 1981; $i$ = 80\arcdeg\ for M82, Shen \& Lo 1995). If we
correct the \ion{C}{1} column densities to face-on values, we obtain
$N$(C) = $9 \times 10^{17}$ cm$^{-2}$ for M83, $1.5 \times 10^{18}$
cm$^{-2}$ for NGC 253, and $3.3 \times 10^{17}$ cm$^{-2}$ for M82. This
result suggests that \ion{C}{1} column densities in M83 measured
perpendicular to the disk may be comparable to those of other starburst
galaxies.

Another way to correct for the differences in the geometry from one
galaxy to another is to compare column density ratios of two different
species, since both species are likely subject to the same geometry
within a given galaxy. Such column density ratios need to be computed
using beams of comparable size for both lines so that the filling
factor of the emission is the same for each line. Since the
\twcoft\ and \ci\ data have approximately the same beam size
(12\arcsec), we will use them to calculate the ratio $N$(C)/$N$(CO).

For the \twcoft\ transition, the column density is given by
$$ N({\rm CO}) = 4.3 \times 10^{14} ~\int T_{\rm MB} dv~{\tau\over{1-e^{-\tau}}}
~~~({\rm cm^{-2}})$$
(White \& Sandell 1995) where $\tau$ is the optical depth of the
\twcoft\ emission and we assume $T_{\rm ex}$ = 23 K. For optically thin
gas, the strength of the CO \jft\ line in the emission peak at
(+5\arcsec,+5\arcsec) indicates $N$(CO) = $1.5 \times 10^{17}$
cm$^{-2}$. This column density is much lower than those previously
found in other starburst galaxies (\eg\ $N$(CO) = $5 \times 10^{18}$
cm$^{-2}$ for M82, G\"usten \etal\ 1993; $N$(CO) $\sim3 \times
10^{19}$ cm$^{-2}$ for NGC 253, Israel \etal\ 1995). One possible
explanation for this low CO column density is that our assumption that
the gas is optically thin is incorrect. We would require an optical
depth $\tau \sim 30$ to obtain CO column densities similar to those in
the starburst galaxy M82. Unfortunately, we do not have \thcoft\ data
with which to estimate the optical depth directly.

Another way to obtain the CO column density is by modeling the CO line
ratios. The \twcoft\ data were convolved to the same beam size as the
\twcott\ data using the data reduction package COMB. The integrated
intensity line ratios for the nucleus of M83 are shown in Figure
\ref{CO43ratios}. We estimate a total uncertainty in the line ratios of
30\%. We used the line ratios for the emission peak at ($+$5\arcsec,
$+$5\arcsec) to perform a Large Velocity Gradient (LVG) analysis. If we
adopt a kinetic temperature of 30 K, \twcott/\jto\ = 1.1 (Wall
\etal\ 1993), \thcott/\jto\ = 0.2 (Wall \etal\ 1993), \twcoto/\joz\ = 1
(Wiklind \etal\ 1990), and \twcoft/\jtt\ = 1.13 (Figure
\ref{CO43ratios}), we obtain a lower limit on the CO column density of
$N$(CO) $\gtrsim 3 \times 10^{18}$ cm$^{-2}$. This value gives us a
ratio of $N$(C)/$N$(CO) $<$ 0.33, which is similar to the value found
for the starburst galaxy NGC 253 (Israel \etal\ 1995). We therefore
adopt a $N$(C)/$N$(CO) ratio of 0.33 $\pm$ 0.10 for M83. It should be
noted that this ratio is only for the emission peak at ($+$5\arcsec,
$+$5\arcsec), while the values for NGC 253 and M82 are averaged over
the nuclear region.

Since \ci\ is thought to form in hot PDR regions associated with star
formation, it is interesting to compare the $N$(C)/$N$(CO) ratios for
these three galaxies to their star formation properties. By using the
H$\alpha$ surface brightness ($\Sigma_{\rm H\alpha}$ in erg s$^{-1}$
pc$^{-2}$) divided by the gas surface density ($\Sigma_{\rm H}$ in
M$_{\odot}$ pc$^{-2}$) as an estimate of the star formation activity in
the inner 20\arcsec\ of each galaxy, Wall \etal\ (1993) find that M82
is the most active (log($\Sigma_{\rm H\alpha}/\Sigma_{\rm H}$) = 34.0)
followed by M83 (log($\Sigma_{\rm H\alpha}/\Sigma_{\rm H}$) = 33.3) and
NGC 253 (log($\Sigma_{\rm H\alpha}/\Sigma_{\rm H}$) = 31.8). In this
analysis, the H$\alpha$ fluxes are not corrected for extinction and so
may not be the best way to measure star formation activity. An
alternate indicator of the strength of the starburst is the
far-infrared luminosity, which indicates that M82 and NGC 253 are
stronger starbursts ($L_{\rm FIR} = 3 \times 10^{10}$ L$_{\odot}$ and
$1.5 \times 10^{10}$ L$_{\odot}$ respectively) than M83 ($L_{\rm FIR} =
4 \times 10^{9}$ L$_{\odot}$) (Telesco \& Harper 1980; Wall
\etal\ 1993). Thus, in our small sample (see a summary in Table
\ref{props}), there is no real indication that a strong starburst
increases the $N$(C)/$N$(CO) ratio.

\subsection{Integrated Intensity Line Ratios}

The physical conditions in a molecular cloud or distant galaxy can be
constrained using integrated intensity line ratios. By assuming the
same regions in space are emitting the observed radiation, dividing the
measured integrated intensities of two emission lines eliminates the
unknown filling factor of the molecular gas in the beam, as long as the
beam sizes are the same. Knowledge of how these line ratios vary in
different conditions (such as cold dark clouds or hot star forming
regions) as well as radiative transfer models allows determination of
the conditions required to produce the observed line ratios.

The \twcoft/\twcott\ integrated intensity line ratios vary
substantially over the central region of M83 (Figure \ref{CO43ratios})
with the highest ratios occurring towards the edges of the emission
peaks. We would expect the \twcoft\ emission to be enhanced near the
hotter regions of star formation in the centre of the galaxy. The
\twcoft/\jtt\ line ratio is highest to the right and left of the upper
and lower peak of Figure \ref{32map}, respectively, and a ridge of high
ratios seems to run from the upper right to the lower left of Figure
\ref{CO43ratios}. This ridge corresponds with the arc of active star
forming regions found by near infrared observations by Gallais
\etal\ (1991; see also Figure 4 of Sofue \& Wakamatsu 1994) as well as
radio continuum observations by Turner \& Ho (1994). We thus conclude
that the peak \twcoft/\jtt\ line ratios are likely excited by the high
temperatures associated with the arc of active star forming regions
that encircles the nucleus of M83. Assuming local thermodynamic
equilibrium (LTE), the high line ratios ($>$1) indicate that the higher
$J$ transitions of CO are optically thin and are likely the result of
the high temperatures and/or densities associated with star formation.
In the LTE approximation, a \twcoft/\jtt\ line ratio no greater than
1.7 is predicted for optically thin gas at an excitation temperature of
50 K, while the line ratio cannot exceed one if both transitions are
optically thick. These data indicate that the \twcoft\ emission at
($-$5\arcsec, +5\arcsec) and (0\arcsec, $-$10\arcsec) from the nucleus
of M83 is likely optically thin while the emission at (+5\arcsec,
+5\arcsec) may be optically thick as suggested in the previous
section.

Since the \twcoft\ and \ci\ data have the same beam size, we are able
to directly divide the observed integrated intensities to obtain line
ratios. Ratios were only calculated for those regions where the
integrated intensity is three times the rms noise. The results are
shown in Figure \ref{CIratios}. The \ci/\twcoft\ integrated intensity
line ratios are uniform at the 2$\sigma$ level and have an average
value of 0.25 $\pm$ 0.03. This value is similar to that found in the
starburst galaxy NGC 253 (0.27, Israel \etal\ 1995). In order to
compare our value to the \ci/\twcooz\ ratio, which is the most commonly
published value, we must scale down our \twcoft\ data to equivalent
\twcooz\ line strengths. We will again focus on the peak of the
\ci\ emission at (+5\arcsec, +5\arcsec). Taking the \ci/\twcoft\ line
ratio of 0.21 (see Figure \ref{CIratios}) and using the
\twcoft/\jtt\ line ratio of 1.02 (see Figure \ref{CO43ratios}), the
\twcott/\jto\ ratio of 1.1 (Wall \etal\ 1993), and the
\twcoto/\joz\ ratio of 1.0 (Wiklind \etal\ 1990) we obtain a
\ci/\twcooz\ ratio of 0.24 with an uncertainty of at least 30\%. This
ratio is higher than the value found in the starburst galaxy M82 (0.11,
Schilke \etal\ 1993), the non-starburst spiral galaxy M33 (0.04-0.18,
Wilson 1997), the starburst galaxy IC 342 (B\"uttegenbach \etal\ 1992),
and our own Milky Way (0.16, Wright \etal\ 1991, assuming
\twcoto/\joz\ = 0.7, Sakamoto \etal\ 1994), but it is lower than that
of the starburst galaxy NGC 253 (Israel \etal\ 1995). This comparison
suggests that there is more \ion{C}{1} being produced in M83 than in
any of these other galaxies except NGC 253, but without accurate
knowledge of the optical depth of the CO emission, this result is
disputable.

While the actual value of the \ci/CO line ratio may not yet tell us
much about the conditions in the different regions of the galaxy, we
can learn about the relative strengths of the processes creating and
destroying \ion{C}{1} and CO by comparing the \ci/CO line ratio over
different regions of the galaxy. If we can neglect optical depth
effects, the \ci/CO ratio is a measure of the relative quantities of
the emitting material. It is interesting that the \ci/\twcoft\ line
ratio is uniform at the 2$\sigma$ level, while the \twcoft/\jtt\ ratio
varies substantially (up to $\sim50\sigma$, see Figure
\ref{CO43ratios}). While we do not know exactly what conditions are
required to produce high concentrations of \ion{C}{1}, the uniformity
of the \ci/\twcoft\ line ratios likely indicates that the processes
creating and destroying CO in the $J$ = 4 state and \ion{C}{1} are
comparable throughout the nucleus of M83. Line strengths are
simultaneously dependent on temperature and optical depth, which
suggests that the similarity in the \ci/\twcoft\ line ratios can be
explained most simply by the emission originating in regions containing
similar physical conditions. This interpretation suggests that the
\twcoft\ emission may originate in the hot photodissociation regions
surrounding massive stars that form inside the nuclear starburst of
M83, since these are the regions that are likely to form atomic
carbon.

\section{Conclusions \label{concl}}

This paper presents \ci, \twcoft, and \twcott\ maps of the nucleus of the
barred spiral galaxy M83 taken at the JCMT. The main results are summarized
below. 

\begin{enumerate}

\item{We observe a double peaked structure in the molecular emission
consistent with gas inflow along the bar collecting at the inner
Lindblad resonance. The \twcoft\ emission suggests that some of the
molecular gas has made it into the nucleus and is being heated by and
possibly fueling the nuclear starburst. This result indicates that
nuclear starbursts may occur even in galaxies which exhibit a double
peaked emission structure, in contrast to the findings of Kenney
\etal\ (1992).}

\item{We observe different morphologies in the \twcoft\ channel maps
than in the \twcott\ and \twcooz\ channel maps. These data suggest that
\twcooz\ emission may not always be a good tracer of molecular gas in
starburst galaxies, as the CO may be heated sufficiently to produce
little emission in the \joz\ line. Thus, discretion should be applied in
the interpretation of \twcooz\ emission as a tracer of molecular gas in
starburst regions.}

\item{The observations also suggest that the double peaked emission may
be the result of a molecular ring out of the plane of the galaxy
oriented nearly perpendicular to the main disk. This torus of cooler
gas would need to contain a disk of hotter gas that fills its central
void in order to explain the observed morphology.}

\item{The \ci\ line strength indicates carbon column densities of $1
\times 10^{18}$ cm$^{-2}$ while CO emission indicates CO column
densities of $\sim 3 \times 10^{18}$ cm$^{-2}$ and H$_2$ column
densities of $3 \times 10^{22}$ cm$^{-2}$ at the peak of the emission.
The $N$(C)/$N$(CO) ratio at this location is 0.33 $\pm$ 0.10
which is similar to those found in other starburst
galaxies.}
 
\item{The twelve \ci/\twcoft\ line ratios in the inner 24$''\times
24''$ are uniform at the 2$\sigma$ level and have an average value of
0.25 $\pm$ 0.03, similar to those of other starburst galaxies. The
uniformity of the line ratios suggests that both the high-excitation CO
emission and atomic carbon form in photodissociation regions in the
starburst nucleus.}

\item{The \twcoft/\twcott\ integrated intensity line ratios vary
substantially over the central region of M83 with the highest ratios
occurring towards the edges of the emission peaks. The \twcoft/\jtt\ line
ratios seem to be enhanced along an arc of active star forming regions. The
high line ratios ($>$1) indicate that the higher $J$ transitions of CO are
optically thin and are likely the result of the high temperatures and/or
densities associated with star formation. }

\end{enumerate}

\acknowledgments The JCMT is operated by the Royal Observatories on behalf
of the Particle Physics and Astronomy Research Council of the United
Kingdom, the Netherlands Organization for Scientific Research, and the
National Research Council of Canada. This research has been supported by a
research grant to C. D. W. from NSERC (Canada).

\newpage

\figcaption[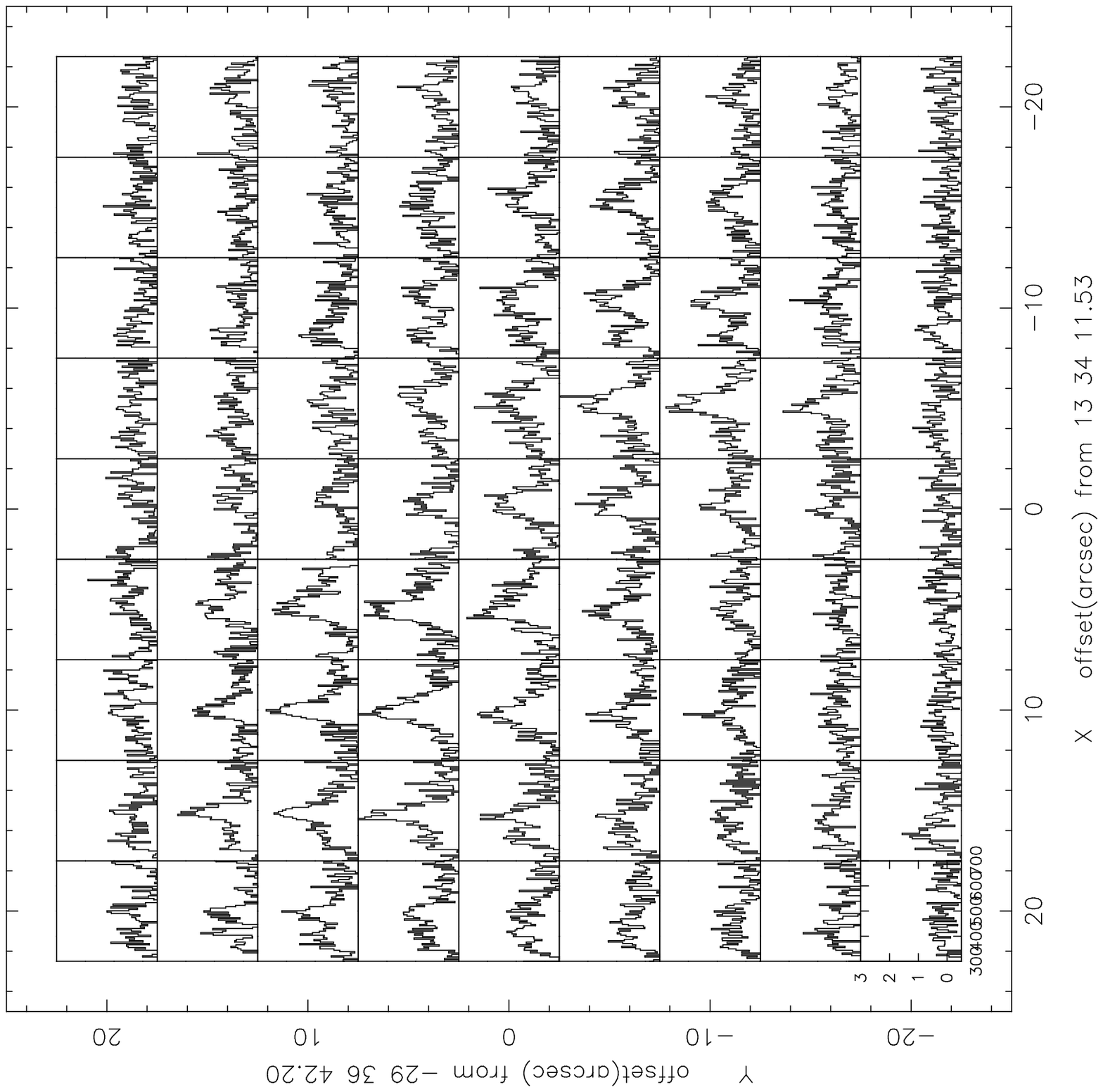]{Individual \twcott\ spectra for the centre
of the galaxy M83. The orientation is such that north corresponds to
an angle 135\arcdeg\ from the positive $x$-axis and the bar runs along
the $x$-axis. The temperature scale is main beam temperature (\tmb).
The map is centred at $\alpha$ = 13$^{\rm h}$34$^{\rm m}$11\fs 53,
$\delta$ = $-$29\arcdeg 36\arcmin 42\farcs 20 (B1950.0). The velocity
range plotted is 300 \kms\ to 700 \kms\ at a resolution of 16.2 \kms\
(18.75 MHz) and the vertical range is $-$0.5 K to 3.0 K.
\label{32grsp}}

\figcaption[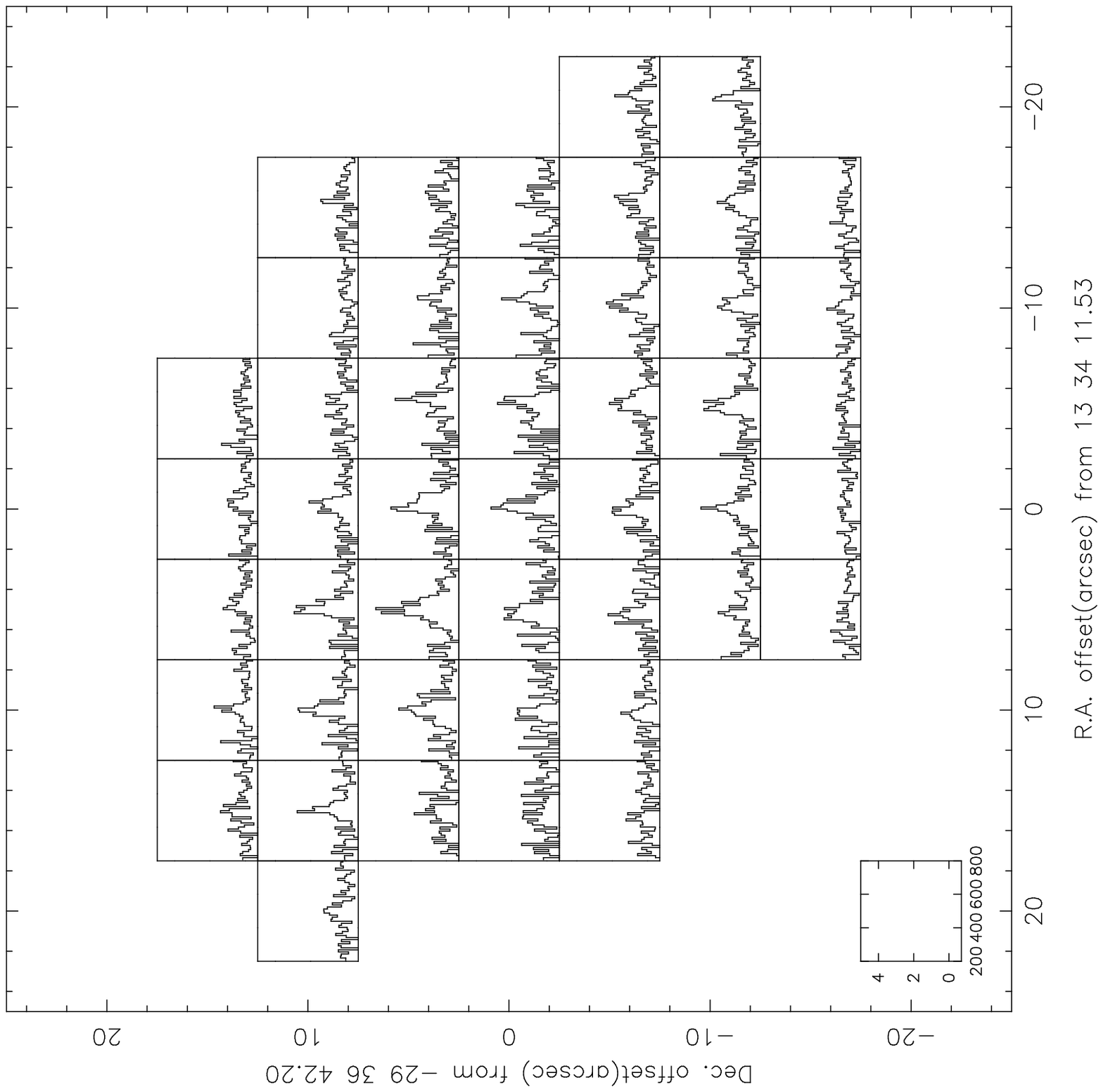]{Individual \twcoft\ spectra for M83. The
temperature scale is main beam temperature (\tmb). The map centre and
orientation is the same as Figure \ref{32grsp}. The velocity range
plotted is 200 \kms\ to 800 \kms\ at a resolution of 12.2 \kms\ (18.75
MHz) and the vertical range is from $-$0.5 K to 5.0 K.
\label{43grsp}}

\figcaption[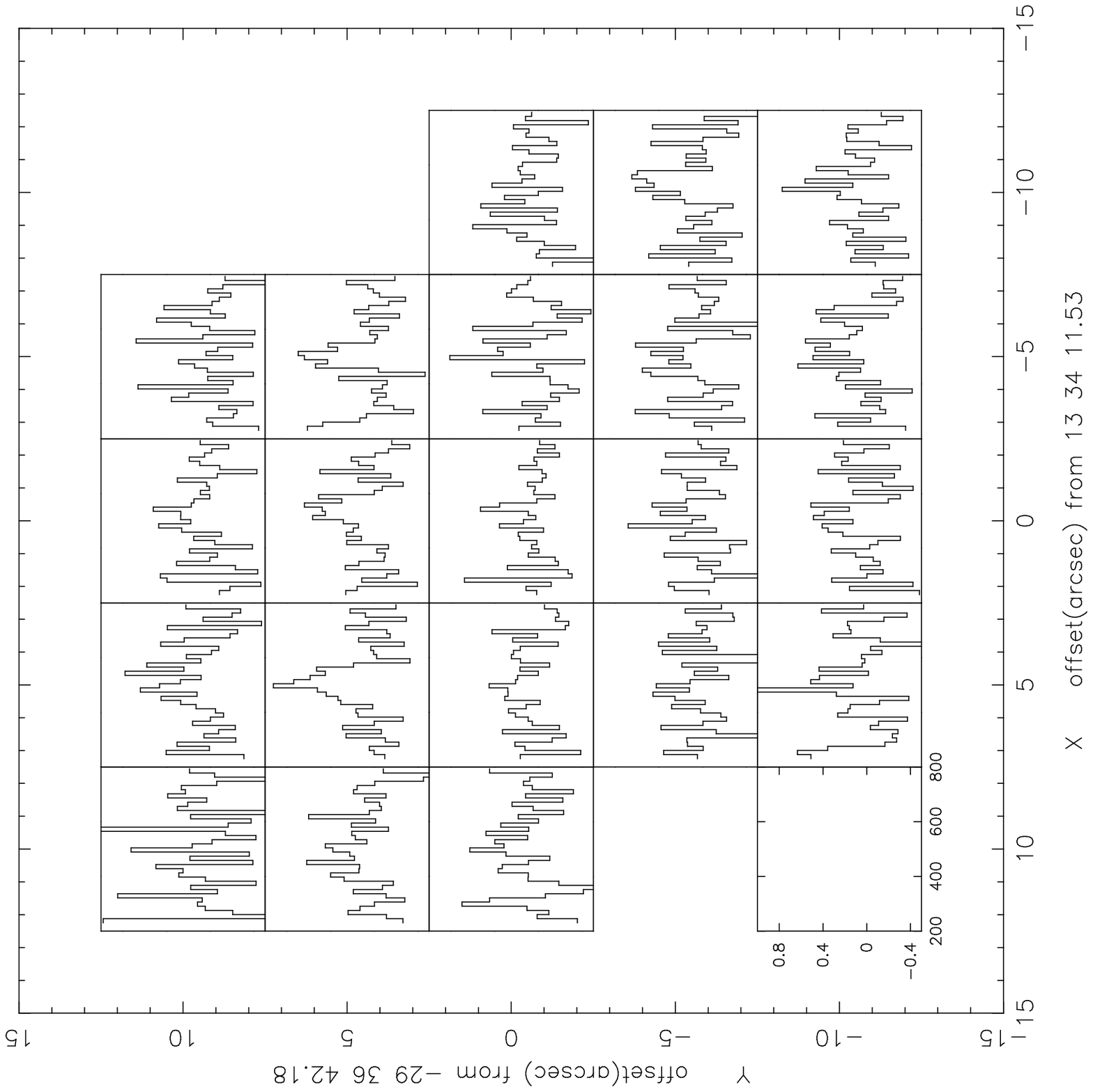]{ Individual \ci\ spectra for M83. The
temperature scale is main beam temperature (\tmb). The map centre and
orientation is the same as Figure \ref{32grsp}. The velocity range
plotted is 200 \kms\ to 800 \kms\ at a resolution of 15.1 \kms\ (25
MHz) and the vertical range is $-$0.5 K to 1.0 K.
\label{CIgrsp}}

\figcaption[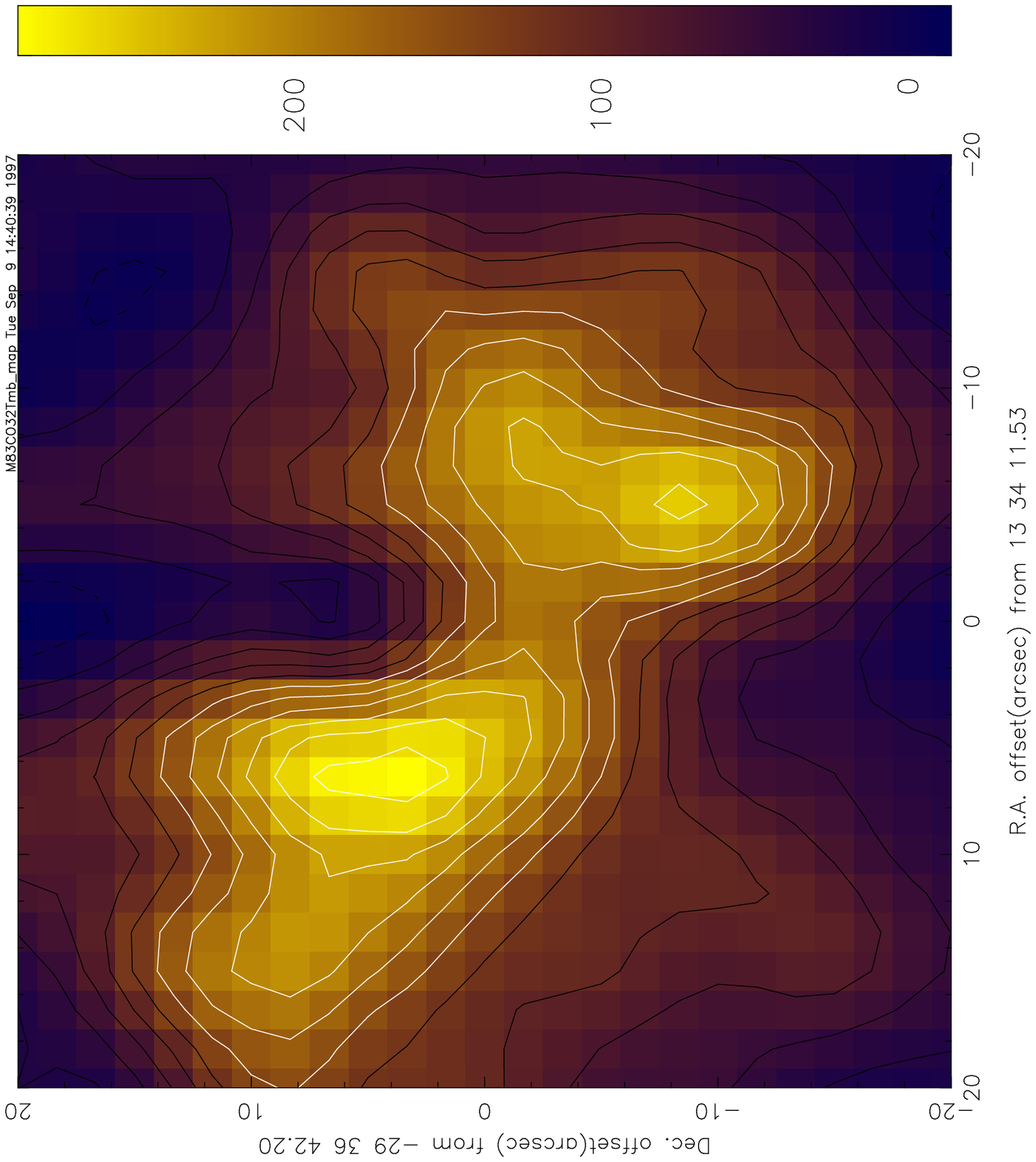]{Integrated intensity map for the \twcott\
transition in M83. The emission is integrated from 400 \kms\ to 600
\kms\ and is given in K \kms\ (\tmb). The contour interval is 25 K
\kms\ with the lowest contour being $-$25 K \kms. The maximum value is
290 K \kms. The emission shows a double peaked structure that is
consistent with gas inflow along the bar collecting at the ILR, or
possibly an inclined ring of molecular gas.
\label{32map}}

\figcaption[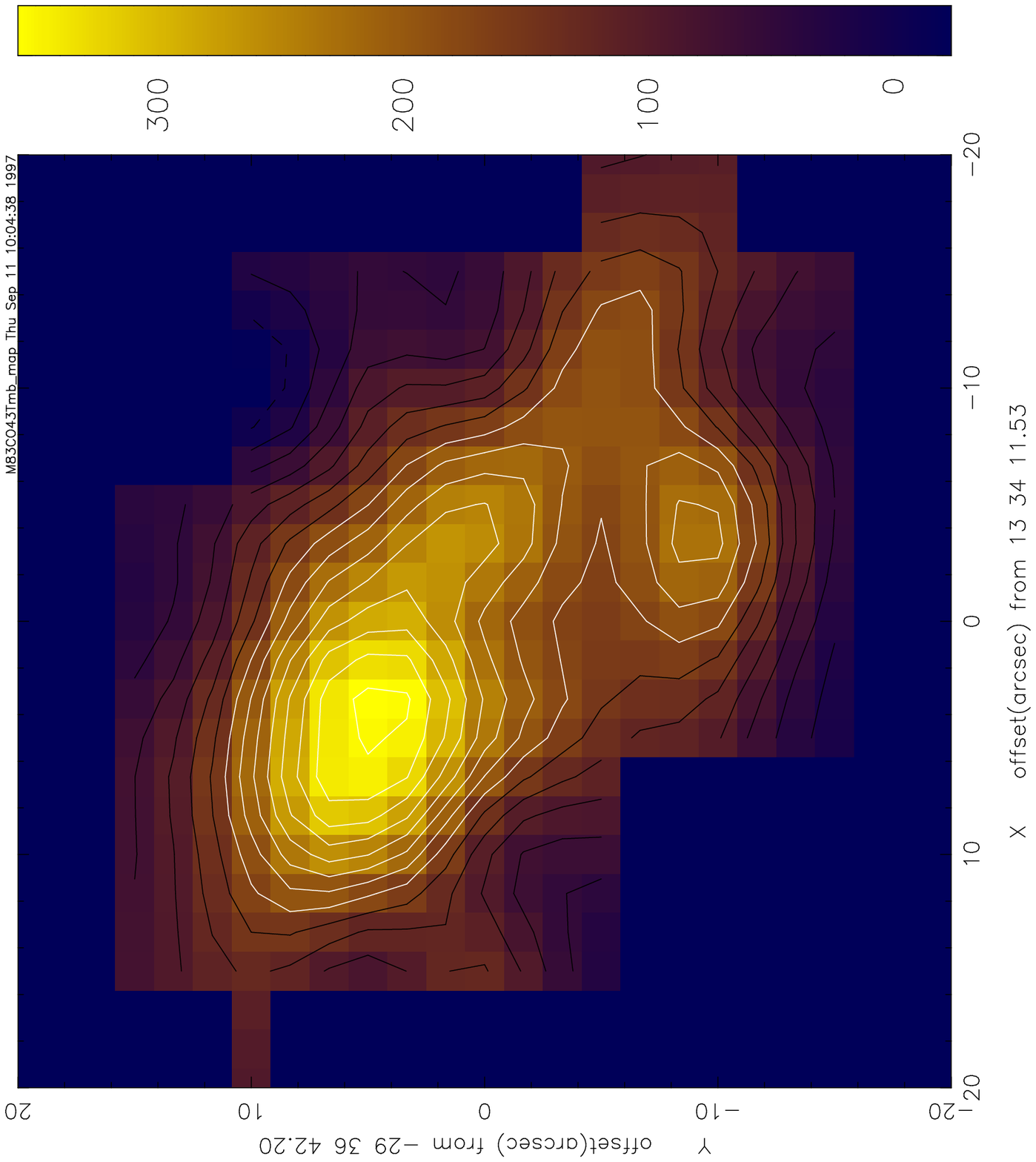]{Integrated intensity map for the \twcoft\
transition in M83. The emission is integrated from 400 \kms\ to 600
\kms\ and is given in K \kms\ (\tmb). The contour interval is 25 K
\kms\ with the lowest contour being $-$25 K \kms. The maximum value is
357 K \kms. These data also show the same double peaked structure
as the \twcott\ data.
\label{43map}}

\figcaption[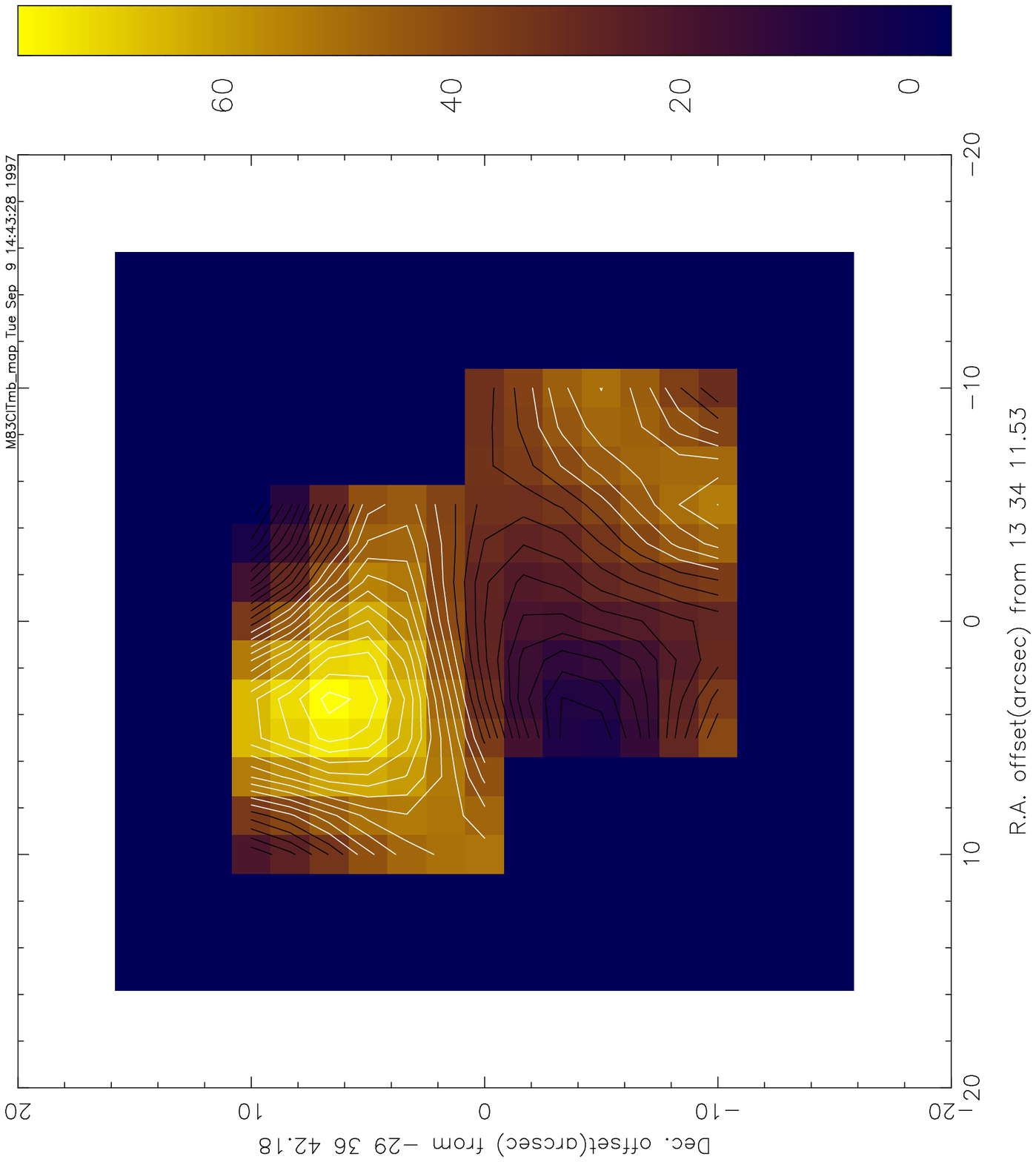]{Integrated intensity map for the
\ci\ transition in M83. The emission is integrated from 400 \kms\ to
600 \kms\ and is given in K \kms\ (\tmb). The contour interval is 5.0 K
\kms\ with the lowest contour being $-$5.0 K \kms. The maximum value is
78 K. These data show a similar `elbow' structure as the
\twcoft\ data. The peaks overlap (within pointing uncertainties) the
two peaks and bend in the \twcoft\ data.
\label{CImap}}

\figcaption[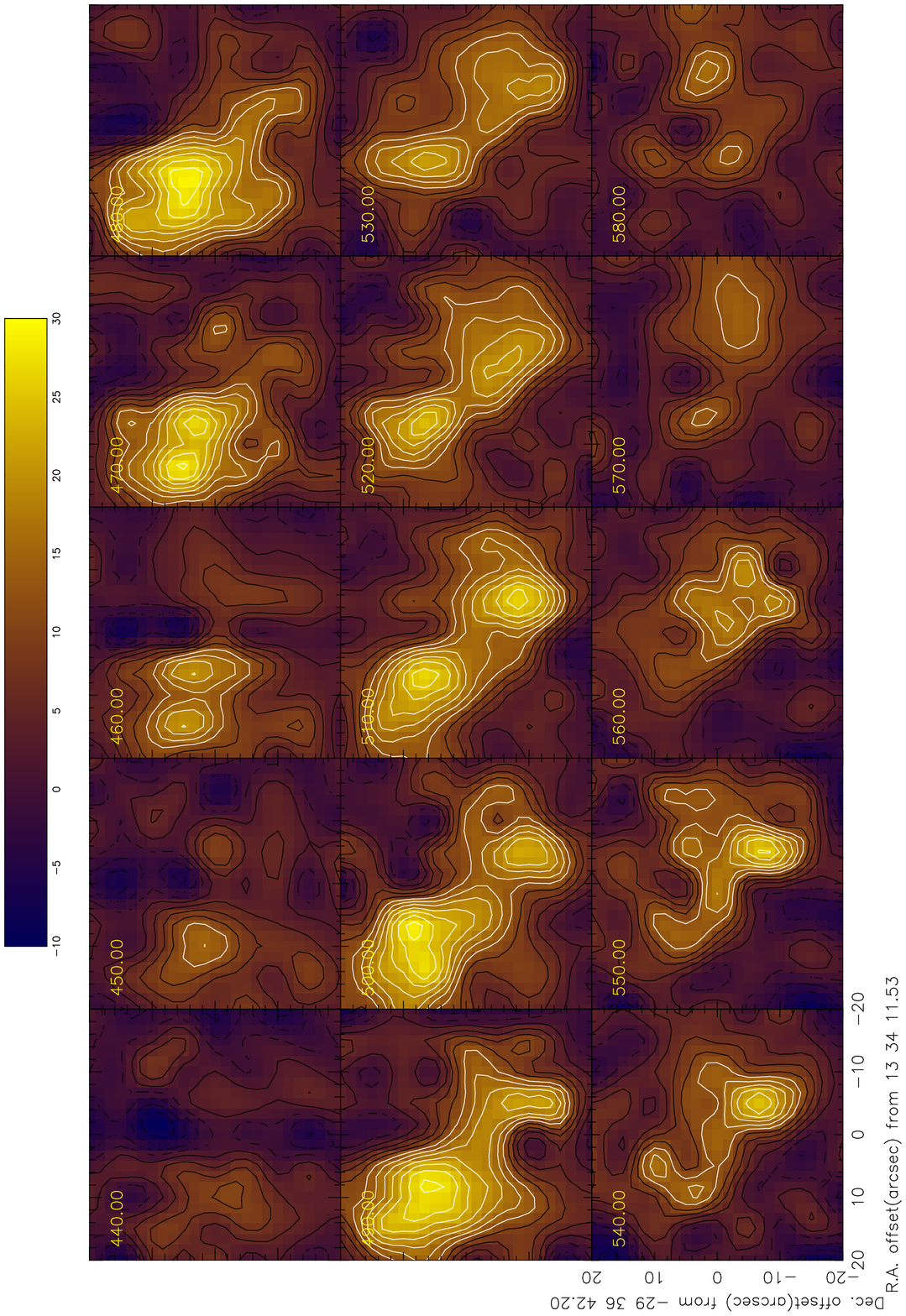]{Integrated intensity channel maps for the
\twcott\ transition in M83. The integrated emission is given in K \kms\
(\tmb). The velocity interval is 10 \kms\ with the first panel
centred at 440 \kms. Contours levels are shown at 2.5 K \kms\ intervals
with the lowest contour at $-$7.5 K \kms. In this figure we can see the
emission to the north peak at velocities lower than the emission to the
south.
\label{32chanmap}}

\figcaption[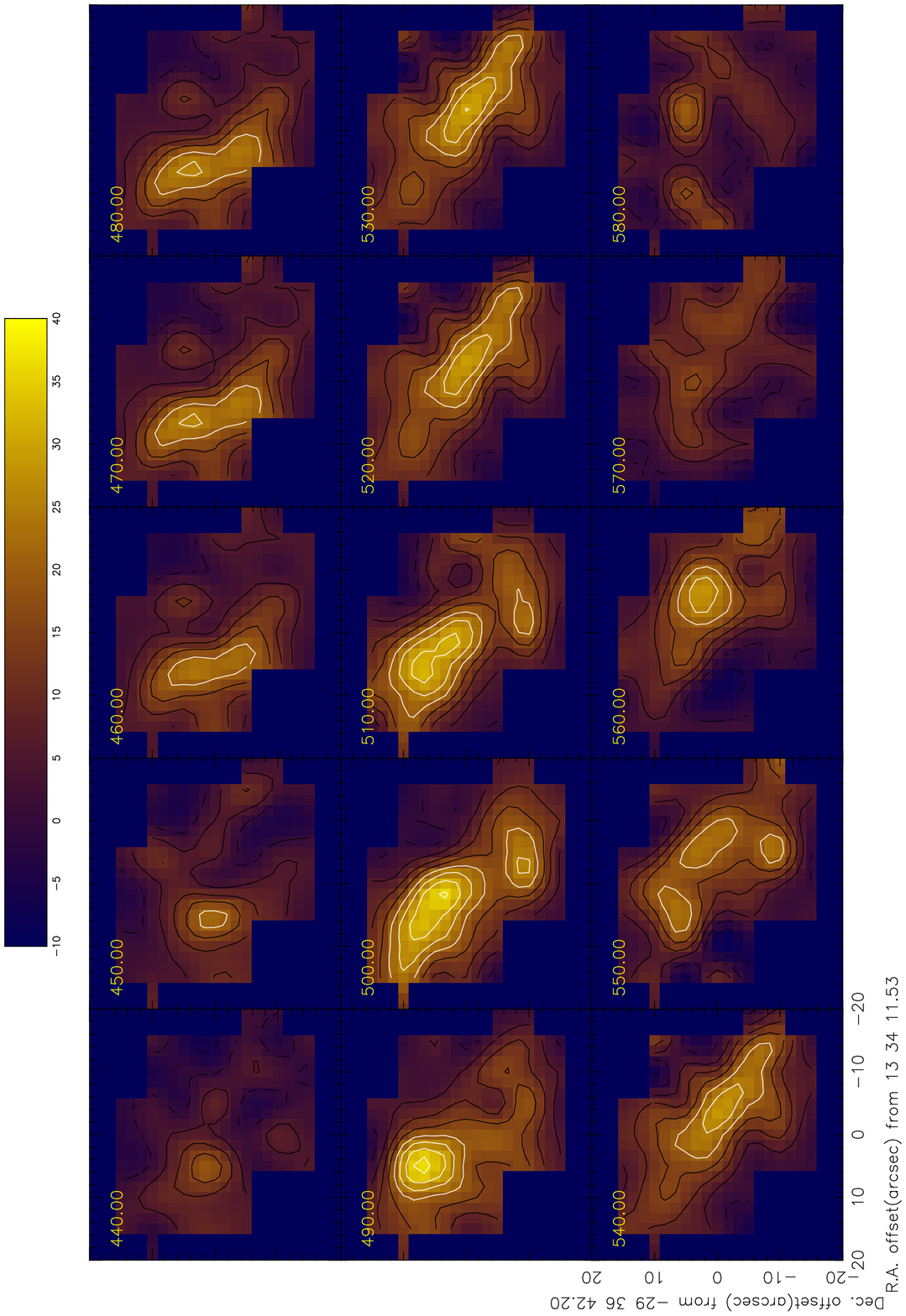]{Integrated intensity channel maps for the
\twcoft\ transition in M83. The integrated emission is given in K \kms\
(\tmb). The velocity interval is 10 \kms\ with the first panel
centred at 440 \kms. Contours levels are shown at 5.0 K \kms\ intervals
with the lowest contour at $-$5.0 K \kms. In this figure we can see the
emission peak shift to the south due to the rotation of the molecular
gas. Note how it peaks at the centre near velocities of $\sim$520 \kms,
in contrast to the \twcott\ data.
\label{43chanmap}}

\figcaption[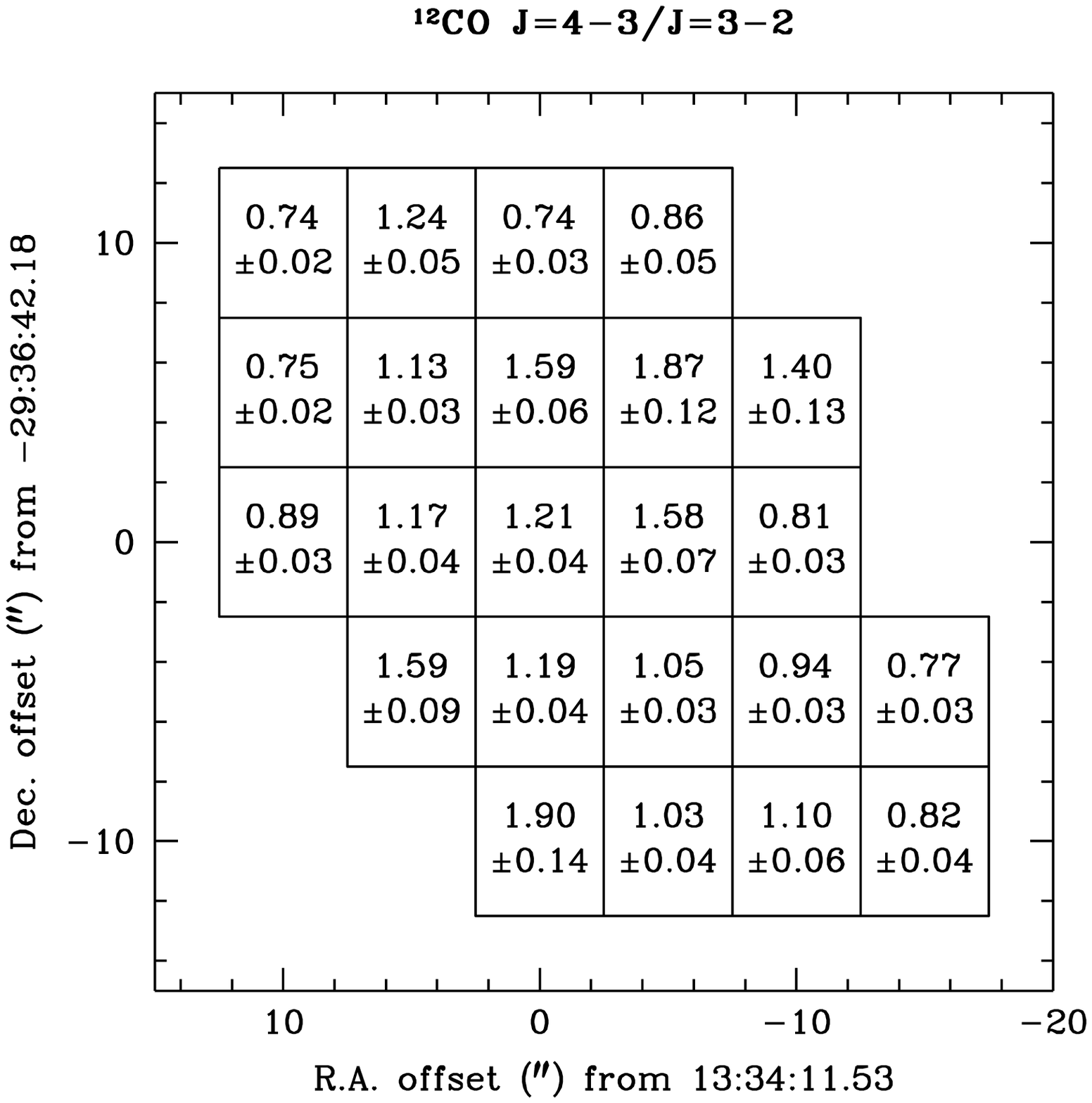]{Integrated intensity line ratios for
\twcoft/\twcott\ at 22 locations in M83. The \twcoft\ data has been
convolved to the same beam size as the \twcott\ data. The line ratio
peaks in a rough line that runs from the upper right to the lower left
corner of the diagram. This strip corresponds to a ring of active star
forming regions.
\label{CO43ratios}}

\figcaption[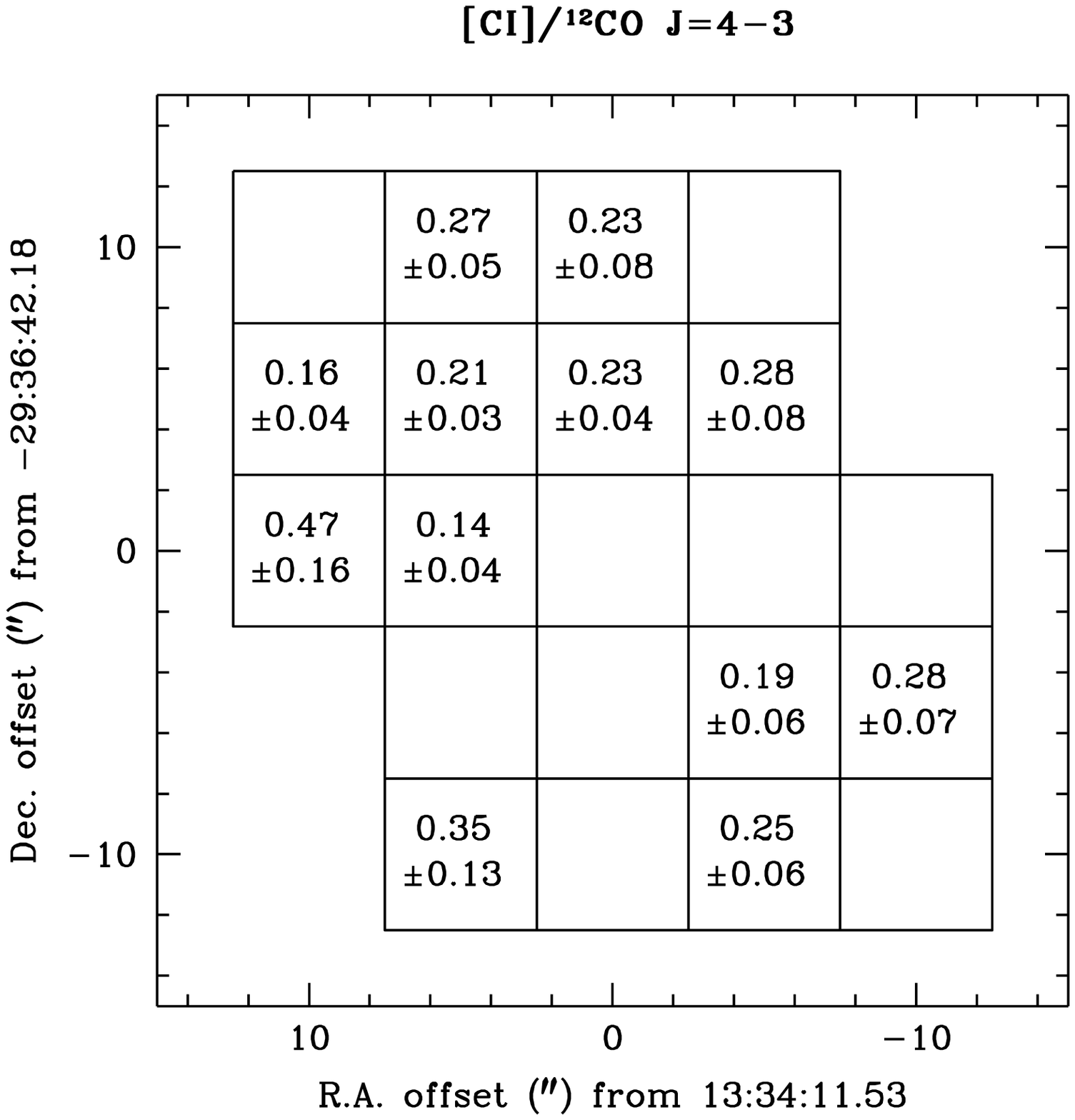]{Integrated intensity (\tmb) line ratios for
\ci/\twcoft\ at 12 locations in M83. Only those locations where the
signal to noise in the \ci\ and \twcoft\ integrated intensity was
greater than three are shown. The data were taken with comparable beam
sizes at the JCMT. The line ratios are uniform at the 2$\sigma$ level,
which indicates that similar processes are responsible for \ion{C}{1}
and \twcoft\ emission from the nucleus of M83.
\label{CIratios}}

\clearpage
{\small
\begin{deluxetable}{lccccccl}
\tablecaption{Properties of Starburst Galaxies \label{props}}
\tablehead{
            \colhead{Galaxy} &
            \colhead{$i$} &
            \colhead{$N$(C)} &
            \colhead{$N$(CO)} &
            \colhead{$N({\rm C})\over{N({\rm CO})}$} &
            \colhead{$L_{\rm FIR}$} &
            \colhead{log($\Sigma$ H$\alpha$/$\Sigma$H)} &
            \colhead{References}  \nl  
            \colhead{~} &
            \colhead{~} &
            \colhead{(cm$^{-2}$)} &
            \colhead{(cm$^{-2}$)} &
            \colhead{~} &
            \colhead{(L$_{\odot}$)} &
            \colhead{log(${\rm erg~ s^{-1}~ pc^{-2}\over{M_{\odot}~ pc^{-2}}}$)} &
            \colhead{~} \nl
           }
\startdata
M83 & 24\arcdeg & $1 \times 10^{18}$ & $> 3 \times 10^{18}$ & 0.33$\pm$0.10  & $4 \times 10^{9}$ &
33.3 & 1, 2, 3 \nl
M82 &  80\arcdeg & $1.9 \times 10^{18}$ & $5 \times 10^{18}$ & 0.5  & $3 \times 10^{10}$ &
34.0 & 2, 3, 4, 5, 6 \nl 
NGC 253 & 78\fdg 5 & $7.7 \times 10^{18}$ & $3 \times 10^{19}$ & 0.2 -- 0.3  & $1.5 \times 10^{10}$ & 31.8 & 2, 3, 7, 8 \nl 
\enddata
\tablenotetext{~}{(1) Comte 1981. (2) Telesco \& Harper 1980. (3) Wall \etal\
1993. (4) Shen \& Lo 1995. (5) White \etal\ 1994. (6) G\"usten \etal\ 1993. (7)
Pence 1981. (8) Israel \etal\ 1995.}

\end{deluxetable}
}

%\clearpage
%\plotone{figure1.ps}
%\newpage
%\plotone{figure2.ps}
%\newpage
%\plotone{figure3.ps}
%\newpage
%\plotone{figure4.ps}
%\newpage
%\plotone{figure5.ps}
%\newpage
%\plotone{figure6.ps}
%\newpage
%\plotone{figure7.ps}
%\newpage
%\plotone{figure8.ps}
%\newpage
%\plotone{figure9.ps}
%\newpage
%\plotone{figure10.ps}


\begin{references}

\reference{ana96} Anantharamaiah, K.~R., \& Goss, W.~M., 1996, \apj, 466, L13

\reference{bal90} Balcells, M., \& Quinn, P.~J., 1990, \apj, 361, 381

\reference{boh83} Bohlin, R.~C., Cornett, R.~H., Hill, J.~K., Smith, A.~M.,
\& Stecher, T.~P., 1983, \apj, 274, L53

\reference{boi90} Boisse, P., 1990, \aap, 228, 483

\reference{but92} B\"uttgenbach, T.~H., Keene, J., Phillips, T.~G., \&
Walker, C.~K., 1992, \apj, 397, L15 

\reference{cmb88} Combes, F., 1988, in Galactic and Extragalactic Star
Formation, ed.~R.~E.~Pudritz \& M.~Fich (Dordrecht: Kluwer), 475

\reference{com81} Comte, G., 1981, \aaps, 44, 441

\reference{cow85} Cowen, J.~J., \& Branch, D., 1985, \apj, 293, 400 

\reference{gal91} Gallais, P., Rouan, D., Lacombe, F., Tiphene, D., \&
Vauglin, I., 1991, \aap, 243, 309

\reference{gus93} G\"usten, R., Serabyn, E., Kasemann, C., Schinckel, A.,
Schneider, G., Schultz, A., \& Young, K., 1993, \apj, 402, 537

\reference{han90} Handa, T., Nakai, N., Sofue, Y., Hayashi, M., \&
Fujimoto, M., 1990, \pasj, 42, 1

\reference{han94} Handa, T., Ishizuki, S., \& Kawabe, R., 1994, in
Astronomy with Millimeter and Submillimeter Wave Interferometry, IAU
Colloquium No.~140, ed.~M.~Ishiguro \& W.~J.~Welch, 341

\reference{her95} Hernquist, L., \& Mihos, J.~C., 1995, \apj, 448, 41

\reference{ish90} Ishizuki, S., Kawabe, R., Ishiguro, M., Okumura, S.~K.,
Morita, K.-I., Chikada, Y., \& Kasuga, T., 1990, \nat, 344, 224

\reference{isr95} Israel, F.~P., White, G.~J., \& Baas, F., 1995, \aap,
302, 343

\reference{kee85} Keene, J., Blake, G.~A., Phillips, T.~G., Huggins, P.~J.,
\& Beichman, C.~A., 1985, \apj, 299, 967

\reference{ken92} Kenney, J.~D.~P., Wilson, C.~D., Scoville, N.~Z.,
Devereux, N.~A., \& Young, J.~S., 1992, \apj, 395, L79

\reference{lar94} Larkin, J.~E., Graham, J.~R., Matthews, K., Soifer, B.
T., Beckwith, S., Herbst, T.~M., \& Quillen, A.~C., 1994, \apj, 420, 159

\reference{leu84} Leung, C.~M., Herbst, E., \& Huebner, W.~F., 1984, \apjs, 56, 231

\reference{mih94} Mihos, J.~C., \& Hernquist, L., 1994, \apj, 425, L13 

\reference{mor73} Morton, D.~C., Drake, J.~F., Jenkins, E.~B., Rogerson, J.
B., Spitzer, L., \& York, D.~G.,1973, \apj, 181, L103

\reference{pen81} Pence, W.~D., 1981, \apj, 247, 473

\reference{phi81} Phillips, T.~G., \& Huggins, P.~J., 1981, \apj, 251, 533

\reference{plu94} Plume, R., Jaffe, D.~T., \& Keene, J., 1994, \apj, 425,
L49

\reference{qui93} Quinn, P.~J., Hernquist, L., \& Fullager, D., 1993, \apj, 403, 74

\reference{rog74} Rogstad, D.~H., Lockhart, I.~A., \& Wright, M.~C.~H.,
1974, \apj, 193, 309

\reference{sak94} Sakamoto, S., Hayashi, M., Hasegawa, T., Handa, T., \&
Oka, T., 1994, \apj, 425, 641

\reference{sch93} Schilke, P., Carlstrom, J.~E., Keene, J., \& Phillips, T.
G., 1993, \apj, 417, L67

\reference{she95} Shen, J., \& Lo, K.~Y., 1995, \apj, 445, L99

\reference{shl89} Shlosman, I., Frank, J., \& Begelman, M.~C., 1989, \nat,
338, 45

\reference{sof94} Sofue, Y., \& Wakamatsu, K.~-I., 1994, \aj, 107, 1018

\reference{sor76} S\o rensen, S.~-A., Matsuda, T., \& Fujimoto, M., 1976,
\apss, 43, 491

\reference{stu90} Stutzki, J., \& G\"usten, R., 1990, \apj, 356, 513

\reference{tal79} Talbot, R.~J., Jr., Jensen.~E.~B., \& Dufour, R.~J.,
1979, \apj, 229, 91

\reference{tel80} Telesco, C.~M., \& Harper, D.~A., 1980, \apj, 235, 392

\reference{til91} Tilanus, R.~P.~J., Tacconi, L.~J., Sutton, E.~C., Zhou,
S., Sanders, D.~B., Wynn-Williams, C.~G., Lo, K.~Y., \& Stephens, S.~A.,
1991, \apj, 376, 500

\reference{tul88} Tully, R.~B., 1988, Nearby Galaxies Catalogue,
(Cambridge: Cambridge Press)

\reference{tur94} Turner, J.~L., \& Ho, P.~T.~P., 1994, \apj, 421, 122

\reference{wal93} Wall, W.~F., Jaffe, D.~T., Bash, F.~N., Israel, F.~P.,
Maloney, P.~R., \& Baas, F., 1993, \apj, 414, 98

\reference{whi94} White, G.~J., Ellison, B., Claude, S., Dent, W.~R.~F., \&
Matheson, D.~N., 1994, \aap, 284, L23

\reference{whi95} White, G.~J., \& Sandell, G., 1995, \aap, 299, 179

\reference{wik90} Wiklind, T., Rydbeck, G., Hjalmarson, \AA., \& Bergman,
P., 1990, \aap, 232, L11

\reference{wil97} Wilson, C.~D., 1997, \apj, 487, L49

\reference{wri91} Wright, E.~L., \etal, 1991, \apj, 381, 200


\end{references}
\end{document}